\documentclass[acmsmall]{acmart}

\usepackage{graphicx}
\usepackage{xspace}
\usepackage{xcolor}
\usepackage{amsmath}

\let\digamma\relax

\usepackage{amssymb}
\usepackage{multirow}
\usepackage{listings}

\newcommand{\ours}{NACS\xspace}
\newcommand{\oursgraph}{NACS$_g$\xspace}
\newcommand{\ourspath}{NACS$_p$\xspace}
\newcommand{\oursast}{NACS$_a$\xspace}

\graphicspath{{figs/}}

\begin{document}

\title{Deep Code Search with Naming-Agnostic Contrastive Multi-View Learning}

\author{Jiadong Feng}
\authornote{The first two authors contributed equally.}
\email{jdfeng@stu.xmu.edu.cn}
\affiliation{%
  \institution{Key Laboratory of Multimedia Trusted Perception and Efficient Computing, Ministry of Education of China, Xiamen University}
  \city{Xiamen}
  \state{Fujian}
  \country{China}
}

\author{Wei Li}
\authornotemark[1]
\email{weilileopku@gmail.com}
\affiliation{%
  \institution{School of Electronic and Computer Engineering, Peking University}
  \city{Shenzhen}
  \state{Guangdong}
  \country{China}
}

\author{Suhuang Wu}
\email{wusuhuang@stu.xmu.edu.cn}
\affiliation{%
  \institution{Key Laboratory of Multimedia Trusted Perception and Efficient Computing, Ministry of Education of China, Xiamen University}
  \city{Xiamen}
  \state{Fujian}
  \country{China}
}

\author{Zhao Wei}
\email{zachwei@tencent.com}
\author{Yong Xu}
\email{rogerxu@tencent.com}
\author{Juhong Wang}
\email{julietwang@tencent.com}
\affiliation{%
  \institution{Tencent}
  \city{Shenzhen}
  \country{China}
}

\author{Hui Li}
\authornote{Corresponding Author.}
\email{hui@xmu.edu.cn}
\orcid{0000-0001-9139-3855}
\affiliation{%
  \institution{Key Laboratory of Multimedia Trusted Perception and Efficient Computing, Ministry of Education of China, Xiamen University}
  \city{Xiamen}
  \state{Fujian}
  \country{China}
}

\renewcommand{\shortauthors}{Jiadong et al.}

\begin{abstract}
Software development is a repetitive task, as developers usually reuse or get inspiration from existing implementations. Code search, which refers to the retrieval of relevant code snippets from a codebase according to the developer's intent that has been expressed as a query, has become increasingly important in the software development process. Due to the success of deep learning in various applications, a great number of deep learning based code search approaches have sprung up and achieved promising results. However, developers may not follow the same naming conventions and the same variable may have different variable names in different implementations, bringing a challenge to deep learning based code search methods that rely on explicit variable correspondences to understand source code. To overcome this challenge, we propose a naming-agnostic code search method (NACS) based on contrastive multi-view code representation learning. NACS strips information bound to variable names from Abstract Syntax Tree (AST), the representation of the abstract syntactic structure of source code, and focuses on capturing intrinsic properties solely from AST structures. We use semantic-level and syntax-level augmentation techniques to prepare realistically rational data and adopt contrastive learning to design a graph-view modeling component in NACS to enhance the understanding of code snippets. We further model ASTs in a path view to strengthen the graph-view modeling component through multi-view learning. Extensive experiments show that NACS provides superior code search performance compared to baselines and NACS can be adapted to help existing code search methods overcome the impact of different naming conventions. Our implementation is available at \url{https://github.com/KDEGroup/NACS}.
\end{abstract}


\maketitle

\section{Introduction}
\label{sec:intro}

Code search takes search queries that manifest the software developer' intent as inputs and returns desired code snippets.
To enhance software development productivity and quality, programmers use code search tools to find high-quality code snippets in existing projects when they debug, write code, look for code to reuse, or learn API usage~\citep{LiuXLGYG22}.
Existing studies show during software development, roughly one-fifth of the development time is used to search code examples~\citep{BrandtGLDK09,SunFCTHZ22}, indicating that code search has become an essential part of software development.

Due to the pivotal role of code search in software development, much effort has been devoted to improving the quality of code search.
Early works mainly use keyword matching between code snippets and search queries written in natural languages~\citep{LiuXLGYG22}.
These methods typically employ traditional information retrieval algorithms~\citep{McMillanGPXF11,LvZLWZZ15,BajracharyaOL10} to measure the relevance between queries and candidate code snippets.
Recently, the success of deep learning techniques has greatly promoted the development of code search approaches~\citep{LiuXLGYG22,YangXLG21,abs-2311-14901}.
Various deep learning techniques have been introduced to improve code search, 
including but not limited to deep neural networks~\citep{GuZ018}, meta learning~\citep{abs-2201-00150}, and pre-training~\citep{FengGTDFGS0LJZ20,GuoRLFT0ZDSFTDC21}.
They encode queries and code snippets into low-dimension spaces and measure their relevance through the similarity (e.g., cosine similarity) between representation vectors.

\begin{figure*}[t]
    \centering
    \includegraphics[width=1\textwidth]{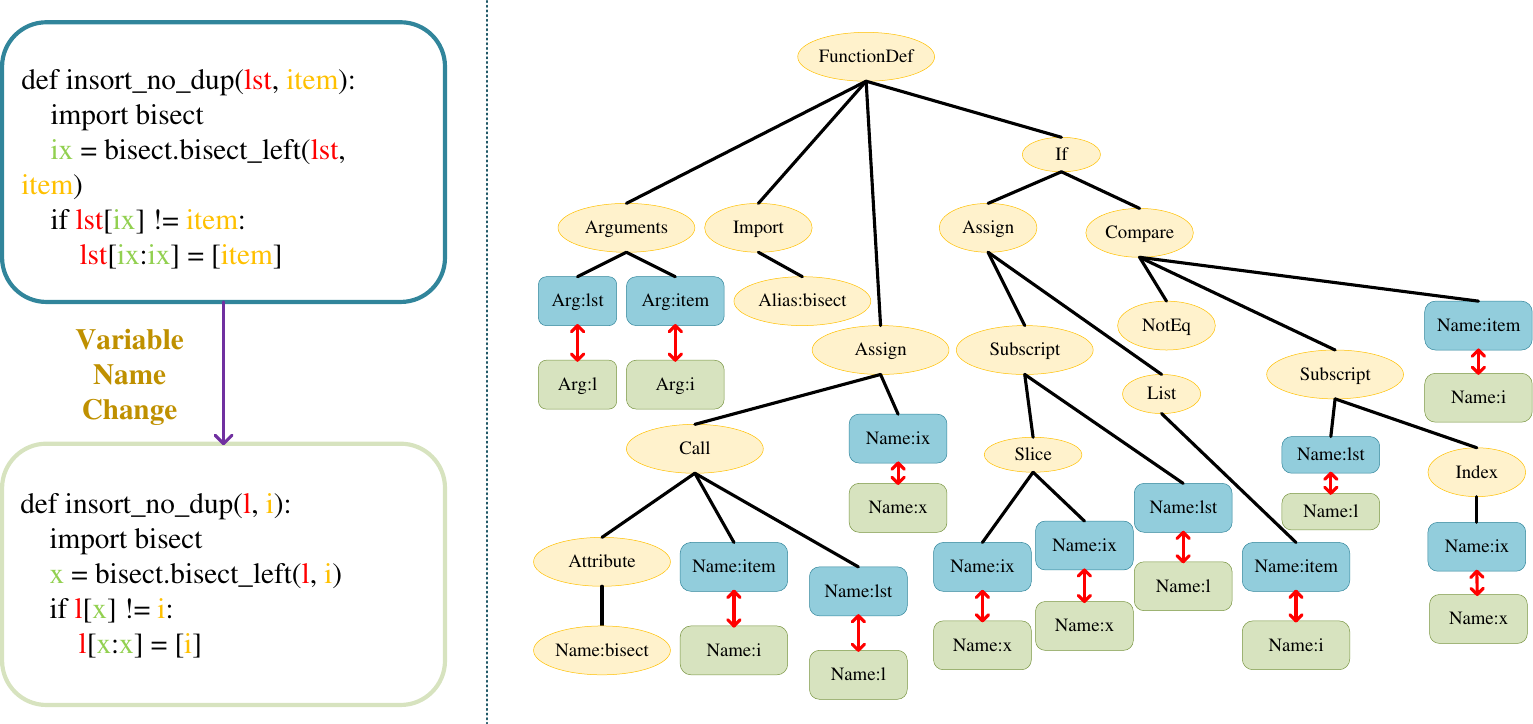}
    \caption{Examples of two code snippets and their corresponding ASTs.}
    \label{fig:ast}
\end{figure*}

However, existing works lack the consideration of the impact of different naming styles (we call it \emph{naming issue} in this paper for simplicity). 
Software developers may not follow the same naming convention, resulting in different naming of the same variable~\citep{GuoRLFT0ZDSFTDC21}.
The naming issue constitutes a barrier for effectively deploying deep learning techniques in code search since (1) it is hard to find correspondences when the same variable has different variable name in different code snippets, and (2) many deep learning techniques rely on entity correspondences (i.e., the same object is always represented by the same representation vector).
In the literature, very few works have been done on remedying the naming issue.
\citet{GuoRLFT0ZDSFTDC21} propose to model the data flow graph, which represents the dependency relation between variables and is the same under different abstract grammars for the same source code, to avoid the impact of different naming conventions in code representation learning.
Nevertheless, source code is different from plain text as it requires strict grammar.
How to avoid the impact of the naming issue when modeling the Abstract Syntax Tree (AST)\footnote{The background of AST is provided in Section~\ref{sec:ast}.}, which represents the abstract syntax of source code and is prevalently modeled in code representation learning~\citep{AllamanisBDS18,LiuXLGYG22,YangXLG21}, is under-exploited.
Previous code representation learning methods not only capture syntactic information contained in ASTs but also encode AST node names, which contain variable names, as the semantic information of ASTs.
Figure~\ref{fig:ast} provides an example to explain why AST-based code representation learning methods suffer from the naming issue. 
The left part of Figure~\ref{fig:ast} provides examples of two python code snippets.
The code snippet at the bottom is the result after changing variable names in the code snippet on the top. 
The right part of Figure~\ref{fig:ast} shows ASTs of the two code snippets.
The blue blocks correspond to the original variable names, and they are replaced by the green blocks after changing variable names.
It is difficult for contemporary code representation learning techniques to understand that corresponding variables in the two code snippets (e.g., lst and l) are the same due to their different names~\citep{GuoRLFT0ZDSFTDC21}.

To overcome the challenge brought by the naming issue, in this paper, we propose a \underline{n}aming-\underline{a}gnostic \underline{c}ode \underline{s}earch method (\ours) based on contrastive multi-view code representation learning.
\ours adopts the idea of contrastive learning~\citep{abs-2006-08218} and multi-view learning~\citep{LiYZ19} to handle the impact of different naming conventions.
In summary, our contributions are:
\begin{itemize}
\item To our best knowledge, we are the first to study how to handle the impact of different naming conventions when modeling ASTs for the code search task.

\item To alleviate the reliance on the explicit correspondence of the same variable in different code snippets in code representation learning, we strip information bound to variable names from input ASTs and focus on capturing intrinsic properties solely from ASTs' structures. Then, we design different semantic-level and syntax-level augmentation techniques to prepare realistically rational data and adopt contrastive learning to design a graph-view modeling component in \ours to enhance the understanding of code snippets.

\item To avoid the information loss caused by ignoring explicit variable correspondences, we design another path-view modeling component that models AST paths. Path-view modeling is used to strengthen the graph-view modeling component through multi-view learning.

\item We conduct experiments to illustrate the effectiveness of \ours on the code search task. We show that (1) The overall performance of \ours is superior to existing code search methods; (2) Existing code search methods perform poorly when facing the naming issue, while the performance of \ours is not affected; (3) The idea of \ours can be adapted to help existing code search methods handle the naming issue in code representation learning.

\end{itemize}

\section{Related Work}
\label{sec:releated work}

In this section, we will illustrate the related work of \ours. 

\subsection{Code Representation Learning}
Software is ubiquitous in modern society and there is an ongoing demand for developing new software and enhancing existing software~\citep{YangXLG21}.
Considering the importance of software development, a great number of assistance tools (e.g., code completion tool~\citep{SvyatkovskiyZFS19,WangL21a}, code search tool~\citep{LvZLWZZ15,abs-2311-14901}, code summarization tool~\citep{FowkesCRALS16,LinOZCLW21}, issue-commit recovery tool~\citep{xZhangWWXWLJ23} and code refactoring tool~\citep{LiuWWXWLJ23}) have been developed to boost the efficiency of software development.
Code and its affiliated documents (e.g., code comments) are important elements in software development.
Strengthening the understanding of code can improve assistance tools and help engineers construct better software~\citep{LeCB20}.
Inspired by the success of deep learning techniques, researchers have recently explored applying deep neural networks (DNNs) to enhance code representation learning and these works can be roughly classified into two categories.

One direction is to specifically design a DNN architecture to model programs.
\citet{Ben-NunJH18} show that a single Recurrent Neural Network (RNN) over the Intermediate Representation of the source code can outperform previous code comprehension approaches.
\citet{ZhangWZ0WL19} split each large AST into a sequence of small statement trees which are encoded by Recursive Neural Network. Based on the sequence of statement vectors, a bidirectional RNN model is used to produce the vector representation of a code fragment. 
\citet{AlonBLY19,AlonZLY19} and \citet{PengZLKHL21} model a code snippet as the set of compositional paths in its AST and adopt RNN or Transformer to learn code representations.
\citet{AllamanisBK18}, \citet{BrockschmidtAGP19} and \citet{CvitkovicSA19} propose to use graphs to represent both the syntactic and semantic structure of code. Then, Graph Neural Network (GNN) is applied for code representation learning over graphs.
\citet{abs-1910-12306} and \citet{BuiYJ21} propose TreeCaps that fuses capsule networks with tree-based convolutional neural networks to enhance code comprehension.
\citet{abs-2203-14285} adopt contrastive learning and GNN to better learn the hierarchical relationships between AST nodes so that the AST hierarchy can be modeled in code representation learning.

Another direction, which recently attracts great attention, is to pre-train Bert-style~\citep{DevlinCLT19}, large-scale models over massive unlabeled code data in a self-supervised manner to gain a better understanding of programs.
CodeBERT~\citep{FengGTDFGS0LJZ20} is a bimodal pre-trained model for natural language (NL) and programming language (PL). It treats source code as plain text and uses a hybrid objective function including standard masked language modeling~\citep{DevlinCLT19} and replaced token detection~\citep{ClarkLLM20} in the pre-training.
CBERT~\citep{abs-2006-12641} is a Bert-based pre-training model for C language. It uses masked language modeling and whole-word masking (i.e., mask all tokens with the same string type) as pre-training objectives.
GraphCodeBERT~\citep{GuoRLFT0ZDSFTDC21} captures data flow (semantic-level structure) instead of syntactic-level structure (e.g., AST) of source code in the pre-training stage. In addition to the standard masked language modeling task, GraphCodeBERT further adopts two structure-aware pre-training tasks: one is to predict code structure edges, and the other is to align representations between source code and code structure.
UniXcoder~\citep{GuoLDW0022} utilizes mask attention matrices with prefix adapters to control the behavior of the model and leverages cross-modal contents like serialized ASTs and code comments to enhance code representation in pre-training.

\subsection{Code Search}

Code search, which is also called code retrieval in the literature, refers to the retrieval of relevant code snippets from a large codebase according to programmer's intent that has been expressed as a query~\cite{LiuXLGYG22,XieLDZW24}.
Code search lowers the cost of learning the usage of new APIs by providing code examples. 
Code search also reduces programmers' cognitive burden and boosts the efficiency of software development by providing reusable code snippets that can be easily adapted.
Therefore, code search has become increasing important in the software development process and attracted considerable attention.

Early works of code search focus on applying traditional information retrieval or natural language processing methods to retrieve relevant code snippets~\cite{XieLDZW24}.
Sourcerer~\citep{BajracharyaNLDRBL06} is a code search engine based on the text search engine Lucene and it searches for relevant code snippets using TF-IDF and some heuristics designed for source code. 
Portfolio~\citep{McMillanPGXF13} finds relevant code snippets by combining various natural language processing and indexing techniques with PageRank~\citep{BrinP98} and spreading activation~\citep{Crestani97} algorithms.
CodeHow~\cite{LvZLWZZ15} uses API documentation to help identify which API a query is likely to refer to. The text similarity score between the query and the API documentation is computed using Vector Space Model~\citep{SaltonWY75}.
These methods simply treat queries and code snippets as plain texts, which can not capture the semantic relationship between the query and the code. 

With the development of deep learning, more and more recent works try to use DNNs to bridge the semantic gap between programming language in code and natural language in query by embedding code snippets and natural language descriptions into a high-dimensional vector space
\citet{GuZ018} firstly apply RNN in code retrieval and propose DeepCS. 
The relevant score is estimated by measuring cosine similarity between query representation and code representation. DeepCS is trained to reduce the distance of matching code-query pairs while keeping unrelated couples apart. Multiple features are considered in DeepCS, including method name, API sequence, code tokens and code descriptions.
PSCS~\citep{abs-2008-03042} extracts AST paths for code search. 
An AST path in PSCS is the path extracted from the AST by walking from one terminal to another. A path has non-terminal nodes in the middle and terminals at both ends.
PSCS deploys Bi-LSTM to encode AST paths and uses them to represent code snippets in code search.
DeGraphCS~\citep{abs-2103-13020} transfers source code into variable-based flow graphs based on the intermediate representation technique and applies GNN to model the VFG for code search.
\citet{LingWWPMXLWJ21} proposes a deep graph matching method for code search. They represent both query texts and code snippets with the unified graph-structured data. Then, query graph and code graph are encoded by relational GNN~\citep{SchlichtkrullKB18} for computing the relevant score.
\citet{abs-2311-14901} study the bias issue of deep code search and propose a reranking based method to alleviate code search biases.

\subsection{Graph Self-Supervised Learning.}

Graph supervised/semi-supervised learning relies on scarce data labels to supervise model learning.
Differently, graph self-supervised learning (GSSL)~\cite{Wu2022} models graphs through handcrafted auxiliary tasks (pretext tasks) that acquire supervision signals from intrinsic graph properties~\citep{abs-2103-00111}.
Since GSSL alleviates label reliance, it draws more and more attention recently~\citep{abs-2102-10757,abs-2105-07342}.

Existing GSSL methods can be grouped into four types according to the used pretext tasks:
\begin{itemize}

\item Generation based graph SSL approaches take the full graph or a subgraph as the model input, and reconstruct one of the components. Typical pretext tasks include feature generation and structure generation. Feature generation learns to reconstruct the feature information of graphs, e.g., mask node or edge features and then recover the masked features~\citep{YouCWS20,abs-2006-10141,HuLGZLPL20,abs-2006-04131}. Structure generation learns to reconstruct the topological structure information of graphs, e.g., randomly mask edges and then recover edges~\citep{abs-2006-04131,abs-2006-10141}.

\item Auxiliary property based methods adopt traditional graph tasks such as clustering~\citep{YouCWS20,SunLZ20}, graph partitioning~\citep{YouCWS20}, shortest distance prediction~\citep{abs-2003-01604} or node similarity prediction~\citep{abs-2006-04131,JinDW0LT21} as the pretext task. These methods have a similar training paradigm with supervised learning since both of them learn with ``sample-label'' pairs~\citep{abs-2103-00111}. The difference is auxiliary property based graph SSL generates pseudo label without manual labeling.

\item Contrast based methods are based on mutual information (MI) maximization~\citep{HjelmFLGBTB19}. They maximize the MI between related graph instances and minimize the MI between unrelated graph instances~\citep{HjelmFLGBTB19,abs-2103-00111}. Two stages are typically required for contrast based graph SSL. 
In the graph augmentation stage, different graph instances are generated through node feature masking~\citep{abs-2006-10141}, node feature shuffling~\citep{abs-1911-08538}, edge dropping/adding~\citep{YouCSCWS20} or subgraph sampling~\citep{HassaniA20,HuLGZLPL20,JiaoXZ0ZZ20,QiuCDZYDWT20}. 
In the graph contrastive learning stage, pretext tasks are leveraged to maximize the MI between positive instances. Contrast can be conducted at same scale or cross scales. The former discriminates instances in an equal scale (e.g., node versus node)~\citep{QiuCDZYDWT20,YouCSCWS20} while the later discriminates instances across different granularities (e.g., node versus subgraph)~\cite{MavromatisK21,RobinsonCSJ21}.

\item Compared to methods that only consider one pretext task, hybrid methods adopt several pretext tasks to leverage different types of supervisions~\citep{abs-2103-00111,YouCWS20}. These methods optimize various pretext tasks in a manner of multi-task learning.

\end{itemize}

\subsection{Multi-view Learning}

Multi-view representation learning (MVL) learns representations from multi-view data~\citep{LiYZ19,YanHMYY21}.
Multi-view data is prevalent in real-world applications where objects are depicted by multi-modal information.
Different views of the same object usually contain complementary information and thus multi-view representation learning can learn more comprehensive representations than single-view learning methods.

Due to the powerful feature abstraction ability, deep learning techniques have exerted considerable influence over the research of MVL.
Various deep learning techniques have been adapted for MVL:
\begin{itemize}
\item Multi-view CNN models from multiple feature sets with access to multi-view information of the target data to obtain more discriminative common representations. Two types of multi-view CNN exist: one-view-one-net multi-view CNN and multi-view-one-net multi-view CNN. One-view-one-net mechanism adopts one CNN for each view and extracts feature representation of each view separately, then multiple representations are fused~\citep{SuMKL15,FengZZJG18}. Multi-view-one-net mechanism feeds multi-view data into the same neural network to get the overall representation~\citep{DouCYQH17}.

\item Multi-view RNN is designed for modeling multi-view sequential data. For example, \citet{MaoXYWY14a} design a multi-view RNN containing a vision network, a language network and a multi-view network for image captioning. \citet{KarpathyF17} propose a multi-view alignment model based on RNN to narrow the interview relationship between visual and textual data for MVL.

\item Multi-view GNN is tailored for MVL over graph data. \citet{FanWSLLW20} design the one2multi graph auto-encoder that learns node  representations by using content information to reconstruct the graph structure from multiple views. \citet{abs-2005-13607} observe that atoms and bonds affect the chemical properties of a molecule. They exploit both node (atom) and edge (bond) information simultaneously to build an expressive, multi-view GNN.

\item Multi-view auto-encoder endorses auto-encoder with the ability of learning from multi-view data. For instance, \citet{NgiamKKNLN11} design a bimodal auto-encoders to find a reconstruction of both audio and video views by minimizing the reconstruction error of the two input views and reconstructed representation. \citet{FengWL14} propose the correspondence auto-encoder for cross-modal retrieval, which simultaneously learns the shared information of multiple modalities and the specific information in each individual modalities.

\item Multi-view GAN aims to overcome the limitation of the single-view GAN, i.e., the single-pathway GAN may learn incomplete representation. For example, \citet{DonahueKD17} design the bidirectional GAN to train an inference network and a generator jointly, which learns a inverse mapping that projects data back into the latent space. \citet{TianPZZM18} propose a two-pathway GAN to maintain the completeness of the learned embedding space.

\item Multi-view contrastive learning is designed for learning consistent representations among different views even when labels are scarce and the multi-view information is incomplete or inconsistent. \citet{0001GLL0021,LinGLBLP23} propose a novel objective that incorporates representation learning and data recovery into a unified contrastive learning framework from the view of information theory.
To overcome the problems of view inconsistency and instance incompleteness, \citet{Yang00B0023} propose robust multi-view clustering with incomplete information (SURE). SURE is a novel contrastive learning paradigm which uses the available pairs as positives and randomly chooses some cross-view samples as negatives.

\end{itemize}

\section{Our Framework \ours}
\label{sec:method}

In this section, we will illustrate the details of \ours, a novel naming-agnostic contrastive multi-view learning based code search method.
\ours mainly consists of three parts: augment data (Section~\ref{sec:transformation}), pre-train via naming-agnostic contrastive multi-view code representation learning (Section~\ref{sec:pretrain}), and enhance code search (Section~\ref{sec:cs}).

\subsection{Data Augmentation}
\label{sec:transformation}

Data augmentation, an essential step in contrastive learning, creates realistically rational data through applying certain transformation strategies that do not affect the information of original data too much.
Without requiring manual labeling, data augmentation enriches the supervision signals.

We adopt different program transformation (PT) strategies and structure transformation (ST) strategies to generate augmented data for code snippets in \ours.
They are applied over the Abstract Syntax Tree (AST) of a code snippet and help \ours better capture intrinsic code features and distinguish different code snippets without requiring more manual labels.

\subsubsection{Abstract Syntax Tree}
\label{sec:ast}

Any programming language has an explicit context-free grammar, and it can be used to parse source code into an AST which represents the abstract syntactic structure of source code~\citep{Ha2016}.
We start by providing the definition of the AST:
\begin{definition}[Abstract Syntax Tree]
An Abstract Syntax Tree (AST) of a code snippet $c$ is a tuple $\left< \mathcal{NT}, \mathcal{T}, r, f \right>$ where $\mathcal{NT}$ is the non-terminal node set, $\mathcal{T}$ is the terminal node set, $r$ is the root node, and $f:\mathcal{NT} \rightarrow (\mathcal{NT} \cup \mathcal{T})$ is a mapping function that maps a non-terminal node to its child nodes.
\end{definition}
An AST is a tree where each non-leaf node corresponds to a \emph{non-terminal} in the context-free grammar specifying structural information (e.g., \texttt{ForStatement} and \texttt{WhileStatement}). 
Each leaf node corresponds to a \emph{terminal} (e.g., variable names and operators) in the context-free grammar encoding program text. 
An AST can be converted back into source code easily. 
From the AST shown in Figure~\ref{fig:ast}, we can see that each non-leaf node contains a type attribute (e.g., \texttt{Arguments}) and each leaf node contains a type attribute and a value attribute (e.g., \texttt{Name:ix} means that the type is \texttt{Name} and the value is \texttt{ix}). 
AST has been widely adopted in designing software engineering tools~\citep{ZhangWZ0WL19}.
On the one hand, compared to plain source code, AST is abstract and it does not include all details such as the punctuation and delimiters.
On the other hand, AST can describe both the lexical and syntactic structural information of the code snippet, and it provides a structural view of the source code which is pivotal in helping \ours achieve naming-agnostic code search.

\subsubsection{PT and ST Strategies}

\begin{figure}[t]
\centering
\includegraphics[width=0.66\textwidth]{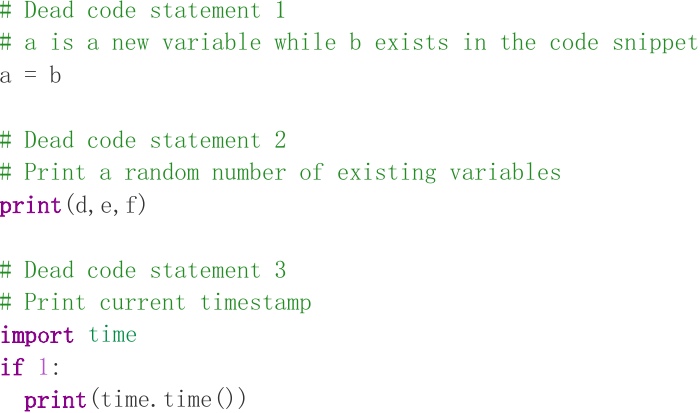}
\caption{Three dead code statements.}
\label{fig:deadcode}
\end{figure}

\begin{figure}[t]
\centering
\includegraphics[width=0.85\textwidth]{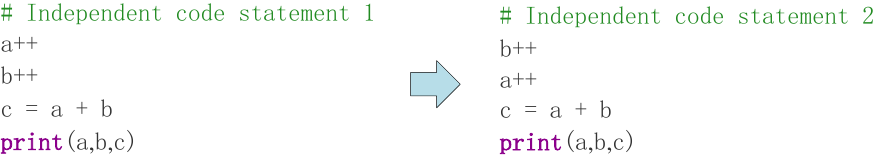}
\caption{Two independent code statements.}
\label{fig:swap}
\end{figure}

\begin{figure}[t]
\centering
\includegraphics[width=0.85\textwidth]{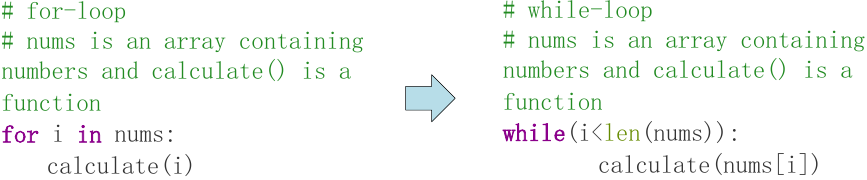}
\caption{An example of two equivalent loops.}
\label{fig:loop}
\end{figure}

PT strategies are semantics-level augmentation while ST strategies are syntax-level augmentation.
We design the following PT and ST strategies for data augmentation in \ours:

\vspace{5pt}
\noindent\textbf{PT Strategies.}
We use three PT strategies to generate augmented versions of a code snippet that do not change the original \emph{semantics} and they can help \ours capture program semantics better:
\begin{itemize}
    \item \textbf{Insert dead code statement}: \ours randomly inserts one dead code statement that does not affect the functionality of the code snippet to a random position of each code snippet. This is equivalent to adding a subtree to the corresponding AST. Figure~\ref{fig:deadcode} illustrates three dead code statements.

    \item \textbf{Swap independent statements}: By analyzing the dependencies of variables, \ours exchanges two statements in each code snippet that do not depend on each other. Figure~\ref{fig:swap} illustrates an example of swapping independent statements. In the example, the order of a++ and b++ does not affect the result and they are independent statements.
    
    \item \textbf{Change loop statements}: By replacing corresponding nodes in the AST, \ours changes for-loop to while-loop or vice versa. If multiple iteration structures exist in one code snippet, one of them will be randomly picked and changed. Figure~\ref{fig:loop} depicts an example of changing for-loop to while-loop. In the example, the for-loop and while-loop are equivalent and changing for-loop to while-loop does not affect the result.
\end{itemize}

\vspace{5pt}
\noindent\textbf{ST Strategies.}
In addition to modeling the semantic meaning of code snippets, capturing the structural features by learning the AST of each code snippet as a graph is commonly adopted in code representation learning~\citep{WangL21a,AllamanisBK18}.
As ASTs represent syntactic meaning of code snippets, structure-level learning is equivalent to capturing \emph{syntactic} information. 
Recently, graph contrastive learning has attracted considerable attention and achieves promising performance in various graph based applications~\citep{abs-2105-07342}.
Nevertheless, existing graph data augmentation strategies are not suitable for code representation learning.
We design two graph transformation methods to generate structure-augmented data of the AST of each code snippet, which enhances the structure-level understanding of \ours:
\begin{itemize}
    \item \textbf{Subtree dropping}: Existing node dropping or edge dropping methods for augmenting graph data~\citep{YouCSCWS20} randomly drop certain portions of vertices and/or edges in the graph. However, doing so will transfer an AST to a disconnected graph and significantly change the structural information of the AST. We opt to randomly discard one subtree of the AST, and the probability of discarding a subtree is inversely proportional to the number of nodes in the subtree.
    
    \item \textbf{Feature shuffling}: We randomly exchange features of nodes in an AST. The motivation of this strategy is that, after the transformation, the new graph is still partially similar to the original AST (i.e., similar in structure but different in node features). The difference between the transformed graph and the original AST graph can help \ours better understand the intrinsic properties of the AST.
\end{itemize}

For a code snippet $c$ of which the AST is denoted by $x_c^q$, three positive samples $x^{p1}_c$, $x^{p2}_c$ and $x^{p3}_c$ are generated in each epoch of training.
$x^{p1}_c$ is constructed by randomly adopting two of the three PT strategies over $c$ together, while $x^{p2}_c$ and $x^{p3}_c$ are generated by adopting two ST strategies individually. 
Since PT strategies require traversing and analyzing ASTs, they are costly and we randomly adopt two of the three PT strategies together for each code snippet.
Differently, the overhead of ST strategies IS relatively smaller. Subtree dropping involves AST traversal without the costly analysis. Feature shuffling does not require AST traversal.
Thus, both ST strategies are adopted on each code snippet, but they are deployed independently to avoid causing a large deviation from the original code snippet.

\subsection{Naming-Agnostic Contrastive Multi-view Code Representation Learning}
\label{sec:pretrain}

\begin{figure*}[t]
    \centering
    \includegraphics[width=1\textwidth]{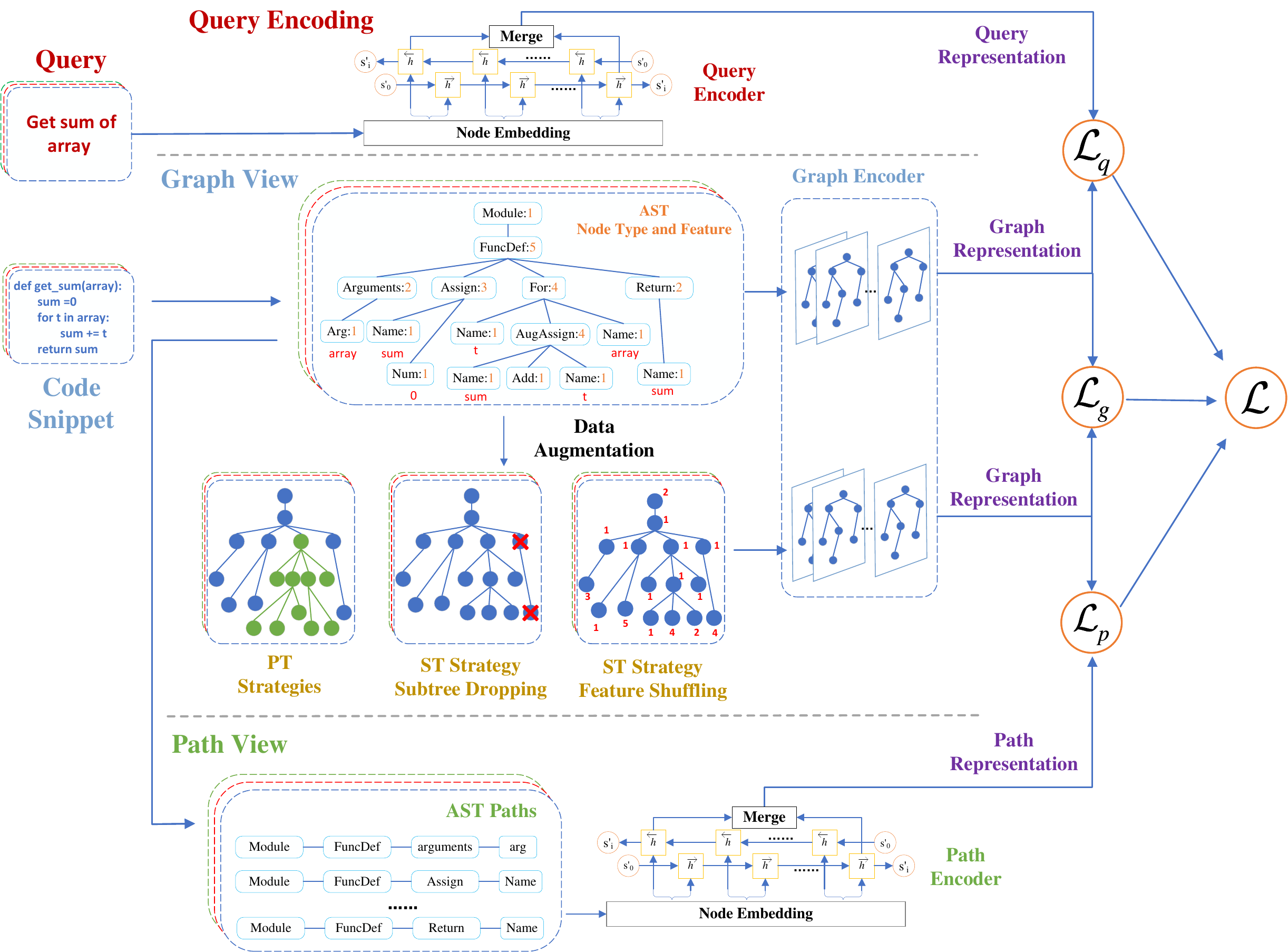}
    \caption{Overview of pre-training \ours via naming-agnostic contrastive multi-view code representation learning.}
    \label{fig:framwork}
\end{figure*}

To overcome the difficulty of aligning variables in different code snippets with same meaning, we design a naming-agnostic contrastive multi-view code representation learning component in \ours.
The core idea is to strip information bound to variable names from input ASTs and focus on capturing intrinsic properties solely from their structures.
To avoid information loss brought by the decoupling, in addition to the common modeling approach that models ASTs as graphs, we model ASTs through multi-view learning over both graph view and path view of ASTs, which helps enhance the understanding of \ours on AST structures.
Furthermore, graph contrastive learning is adopted with the help of augmented data introduced in Section~\ref{sec:transformation} to amplify supervision signals to alleviate the deficiency of manual labels.
Figure~\ref{fig:framwork} provides an overview of pre-training \ours via naming-agnostic contrastive multi-view code representation learning.

In the following, we will illustrate the graph-view modeling component and the path-view modeling component in \ours and how graph contrastive learning is used in each part.

\subsubsection{Graph-View Modeling}

Firstly, \ours captures the syntax-level (structural) information via modeling ASTs as graphs.
This is the \emph{core modeling component} in \ours.
We treat each AST as a graph and pre-train a \underline{g}raph \underline{t}opology encoder (GT-Encoder) to capture the topological information of ASTs (i.e., syntactic information of code snippets).
The core of GT-Encoder is a Graph Neural Network (GNN) and we leverage the Graph Isomorphism Network (GIN)~\citep{XuHLJ19}.
But other GNN architectures can also be adopted.
We adopt the AST graph discrimination task as the pre-training task for graph-view modeling. 
Through contrasting samples, \ours is trained to move an AST graph close to its positive samples but far away from its negative samples in the representation space, helping \ours distinguish AST graphs.
The detailed steps are as follows:
\begin{enumerate}
	\item A mini-batch of $N$ code snippets (and their ASTs) is randomly selected from the codebase. PT and ST strategies are applied on each code snippet to generate its three positive samples. This way, we have a total of $3N$ converted AST graphs.
	
	\item For each node $s$ in an AST graph $g$, we encode its AST node type $\mathbf{b}_{s}^{\text{type}}$ and its degree $\mathbf{b}_{s}^{\text{degree}}$ (i.e., all dimensions of $\mathbf{b}_{s}^{\text{degree}}$ are the degree of $s$ in the AST graph) as part of node features. Additionally, following the initialization method used in existing GNN pre-training approaches~\citep{QiuCDZYDWT20}, we conduct eigen-decomposition on each AST graph's normalized graph Laplacian, $\textit{ s.t. } \mathbf{I}-\mathbf{D}^{-1 / 2} \mathbf{A} \mathbf{D}^{-1 / 2}=\mathbf{U} \mathbf{\Lambda} \mathbf{U}^{\top}$, and then add the top eigenvectors of $\mathbf{U}$ to construct node features of $s$:
    \begin{equation}\label{eq:nodefeature}
        \mathbf{b}_{s} = \mathbf{b}_{s}^{\text{type}} \oplus  \mathbf{b}_{s}^{\text{degree}} \oplus  \mathbf{b}_{s}^{\text{lapla}},
    \end{equation}
    where $\mathbf{b}_{s}$ is node features of $s$, $\mathbf{b}_{s}^{lapla}$ is an eigenvector of $\mathbf{U}$ that corresponds to $s$, ``$\oplus$'' denotes the concatenation operation. Then, each AST is fed to a GIN encoder to generate the node representations:
    \begin{equation}\label{eq:gin}
    \mathbf{h}_{s}^{(k)}=\operatorname{MLP}^{(k)}\big((1+\epsilon^{(k)}) \cdot \mathbf{h}_{s}^{(k-1)}+\sum_{u \in \mathcal{N}(s)} \mathbf{h}_{u}^{(k-1)})
    \end{equation}
    where $\operatorname{MLP}(\cdot)^{(k)}$ denotes the multi-layer percetron for the $k$-$th$ layer, $\mathbf{h}_s^{(k)}$ represents the output for the AST node $s$ at the $k$-$th$ layer of GIN, $\mathcal{N}(s)$ is the set of neighbors of $s$, $\epsilon$ is a learnable parameter, and we set $\mathbf{h}_{s}^{(0)}=\mathbf{b}_{s}$.
    Then, we perform a mean pooling operation on representations of all nodes in $g$ output by GIN. 
    The result is fed to a MLP to get the graph representation $\mathbf{r}_g^{(i)}$ of $g$ from the $i$-$th$ layer. 
    The final graph representation $\textbf{r}_g$ of the AST graph $g$ is the sum of $\{\mathbf{r}_g^{(1)}\cdots\mathbf{r}_g^{(K)}\}$ where $K$ is the number of layers:
    \begin{equation}
        \label{eq:ast_graph_rep}
        \begin{aligned}
        \mathbf{r}_g^{(i)} &= \operatorname{MLP}(\frac{1}{M_g} \sum_{s=1}^{M_g} \mathbf{h}_s^{(i)})\\
        \mathbf{r}_g &= \sum_{i=1}^{K} \mathbf{r}_g^{(i)},
        \end{aligned}
    \end{equation}
    where ${M_g}$ indicates the number of nodes in the AST graph $g$.
    
	\item We separately compute three contrastive losses based on the representation of the AST $x_c^q$ of each code snippet $c$ in the mini-batch and the representation of each positive sample ($x_c^{p1}$, $x_c^{p2}$ and $x_c^{p3}$). Taking $x_c^{p1}$ as an example, the contrastive loss value is defined as follow:
    \begin{equation}\label{eq:graphviewloss1}
    l_g(x_c^q,\,x_c^{p1})\!=\!\log \!\frac{e^{\textit{sim}(\mathbf{r}_{c}^q,\,\mathbf{r}_{c}^{p1}) \!/ \!\tau}}{e^{\textit{sim}\left(\mathbf{r}_{c}^{q},\,\mathbf{r}_{c}^{p1}\right) \!/ \!\tau}\!+\!{\sum_{v=1}^{N}} \!\mathbb{I}_{[v \neq c]}\!(e^{\textit{sim}(\mathbf{r}_{c}^{q},\,\mathbf{r}_{v}^{q})\! /\! \tau}\!+\!e^{\textit{sim}(\mathbf{r}_{c}^{q},\,\mathbf{r}_{v}^{p1}) \!/ \!\tau})},
    \end{equation}
    where $\mathbb{I}_{[v \neq c]}\in \{0,1\}$ is the indicator function that equals 1 if
    $v \neq c$ otherwise 0, $\tau$ is a temperature parameter, and $\textit{sim}(\mathbf{r_1}, \mathbf{r_2})$ is the cosine similarity between $\mathbf{r_1}$ and $\mathbf{r_2}$.
    During optimization, we take other data samples in the same mini-batch as negative samples to avoid the long processing time. 
    The overall graph-view contrastive learning loss in \ours is the sum of the three contrastive losses defined in Equation~\ref{eq:graphviewloss1}:
    \begin{equation}\label{eq:graphviewloss2}
    \mathcal{L}_g=\frac{1}{3N}\sum_{i=1}^{3}\sum_{c=1}^{N}  l_g\left({x}_{c}^{q},\, x_{c}^{p_i}\right).
    \end{equation}
    
\end{enumerate}

\subsubsection{Path-View Modeling}

In addition to the graph view, \ours models AST paths, providing another view of the structural information in ASTs.
Path-view modeling is the \emph{auxiliary modeling component} in \ours and it helps strengthen the understanding of ASTs.

For each AST $x_c^q$ of a code snippet $c$, we extract all its paths from the root node to the leaf node $\mathcal{R} = \{y_{c,1},y_{c,2},\cdots,y_{c,z_c} \}$ where $z_c$ represents the total number of paths in $x_c^q$. 
Since the numbers of paths in different ASTs differ, $z_c$ is not fixed. 
$y_{c,i}=\{n_{c,i,1},n_{c,i,2},\cdots,n_{c,i,l_i}\}$, where $l_i$ represents the path length of the AST path $i$ and $n_{c,i,j}$ ($1\leq j \leq l_i$) is the AST node type of the $j$-th node on the AST path $i$. 
We apply a Bi-LSTM, which can extract information in forward and reverse directions, to encode the AST path.
For an AST path $i$ in $x_c^q$, at each time step $s$, the Bi-LSTM reads the embedding of the $s$-th node in $i$, then computes the hidden states $\mathbf{h}_{c,i,s}$:
\begin{equation}
\label{eq:lstm1}
\begin{aligned}
\overrightarrow{\mathbf{h}}_{c,i,s}&=\overrightarrow{L S T M}\left(e m b\left(n_{c,i,s}\right), \overrightarrow{\mathbf{h}}_{c,i,s-1}\right)\\
\overleftarrow{\mathbf{h}}_{c,i,s}&=\overleftarrow{L S T M}\left(e m b\left(n_{c,i,s}\right), \overleftarrow{\mathbf{h}}_{c,i,s+1}\right)
\end{aligned}
\end{equation}
where $emb(n)$ is the embedding of the AST node type $n$.
Note that, path-view modeling component and graph-view modeling component do not share AST node type representations, i.e., $emb(n)$ is not identical to $\mathbf{b}_{s}^{\text{type}}$ in Equation~\ref{eq:nodefeature}.

We concatenate the last hidden states in forward and backward directions to represent the AST path $i$:
\begin{equation}
\label{eq:lstm2}
\mathbf{h}_{c,i} = \overrightarrow{\mathbf{h}}_{c,i,l_i}\oplus \overleftarrow{\mathbf{h}}_{c,i,1}.
\end{equation}
Then we perform mean pooling on representations $\{\mathbf{h}_{c,1},\cdots,\mathbf{h}_{c,g_c}\}$ of all AST paths in the AST $x_c^q$ to obtain the final path representation $\mathbf{w}_c$ of $x_c^q$:
\begin{equation}
\label{eq:path}
\mathbf{w}_c = \operatorname{mean}\big(\{\mathbf{h}_{c,1},\cdots,\mathbf{h}_{c,g_c}\}\big).
\end{equation}

Taking one positive sample $x_c^{p1}$ out of the three positive samples ($x_c^{p1}$, $x_c^{p2}$ and $x_c^{p3}$) for an AST $x_c^q$ as an example, in path-view modeling, the contrastive loss of the positive pair $\left<x_c^q,\,x_c^{p1}\right>$ is defined as follows:
\begin{equation}
\label{eq:pathviewloss1}
    l_{p}(x_c^q,\,x_c^{p1})\!=\!\log \!\frac{e^{\textit{sim}(\mathbf{r}_{c}^q,\,\mathbf{w}_{c}^{p1}) \!/ \!\tau}}{e^{\textit{sim}\left(\mathbf{r}_{c}^{q},\,\mathbf{w}_{c}^{p1})\right) \!/ \!\tau}\!+\!{ \sum_{v=1}^{N}} \!\mathbb{I}_{[v \neq c]}\!(e^{\textit{sim}(\mathbf{r}_{c}^{q},\,\mathbf{w}_{v}^{q})\! /\! \tau}\!+\!e^{\textit{sim}(\mathbf{r}_{c}^{q},\,\mathbf{w}_{v}^{p1}) \!/ \!\tau})}.
\end{equation}

Equation~\ref{eq:pathviewloss1} is similar to Equation~\ref{eq:graphviewloss1}. 
The difference is that, in Equation~\ref{eq:pathviewloss1}, \ours contrasts representations from graph-view and path-view, i.e., cross-view contrast.
As a comparison, Equation~\ref{eq:graphviewloss1} only contrasts graph-view representations.

The overall path-view loss function is defined as follows:
\begin{equation}\label{eq:pathviewloss2}
\mathcal{L}_p=\frac{1}{3N}\sum_{i=1}^{3}\sum_{c=1}^{N}  l_p\left({x}_{c}^{q},\, x_{c}^{p_i}\right).
\end{equation}

\subsubsection{Query Encoding}

Given a query $b_i=\{w_1, ...,w_t\}$ containing a sequence of $t$ natural language words. 
We use a Bi-LSTM to embed the query into its representation $\textbf{s}_i$.
The encoding procedure is similar to Equation~\ref{eq:lstm1}, Equation~\ref{eq:lstm2} and Equation~\ref{eq:path}.

Then, we adopt the InfoNCE loss~\citep{abs-1807-03748}, a type of contrastive loss function used for self-supervised learning, for modeling queries:
\begin{equation}\label{eq:queryviewloss1}
\mathcal{L}_q=-\frac{1}{M}\sum_{u=1}^{M} \log \frac{e^ {\textit{sim}\left(\mathbf{s}_{u}, \mathbf{r}_{(u)}\right) / \tau}}{\sum_{v=1}^{N} e^{\textit{sim}\left(\mathbf{s}_{u}, \mathbf{r}'_{(v)}\right) / \tau}},
\end{equation}
where $M$ is the number of the training queries, $\mathbf{r}_{(u)}$ is the AST graph representation of the ground-truth code snippet of the query $u$, and $\mathbf{r}'_{(v)}$ is the AST graph representation of the code snippet $v$ in the codebase.
$\mathbf{r}_{(u)}$ and $\mathbf{r}'_{(v)}$ are extracted according to Equation~\ref{eq:ast_graph_rep}.
Note that, for query encoding, we contrast query representations and AST graph representations.
AST graph modeling is the primary modeling component in \ours.
And we do not incorporate path representations (from the path view) in Equation~\ref{eq:queryviewloss1} to reduce the overhead when encoding queries, as AST path encoding involves traversing multiple AST paths for each AST.

\subsubsection{Putting All Together}

The overall objective for optimizing the contrastive multi-view learning component in \ours is defined as:
\begin{equation}
\label{eq:loss}
\mathcal{L}=\mathcal{L}_g + \mathcal{L}_p + \mathcal{L}_q.
\end{equation}

\subsection{Enhance Code Search with \ours}
\label{sec:cs}

Next, we describe how pre-trained \ours can be used to enhance code search.

\subsubsection{Derive Code Representations}
\label{sec:coderep}

For each code snippet $c$ in the codebase, we feed its code token sequence into the pre-trained CodeBERT~\citep{FengGTDFGS0LJZ20} to obtain its code-token representation:
\begin{equation}
\mathbf{h}_{c}^{code} = \textit{CodeBERT}_{code}(c).
\end{equation}
CodeBERT is a bimodal model for natural language and programming language which enables high-quality text and code embeddings to be derived. 

After that, the AST graph $x_c^{q}$ of the code snippet $c$ is fed to the graph encoder (Equation~\ref{eq:ast_graph_rep}) in the contrastive multi-view component of \ours to generate the AST representation of $c$: 
\begin{equation}
\mathbf{h}_{c}^{AST} = \textit{NACS}_{AST}(c).
\end{equation}
Similar to the objective of query encoding defined in Equation~\ref{eq:queryviewloss1}, only the primary modeling method in \ours (i.e, graph-view modeling) is used at the search time to avoid the cost of traversing multiple AST paths so as to make the search phase lightweight.
As the graph encoder is trained together with the path encoder in multi-view learning, it has been endowed with the ability to understand ASTs in multiple views.

In summary, we have a code-token representation $\mathbf{h}_{c}^{code}$ generated by the pre-trained CodeBert and an AST representation $\mathbf{h}_{c}^{AST}$ generated by the pre-trained \ours for each code snippet $c$.

\begin{figure}[t]
    \centering
    \includegraphics[width=0.9\textwidth]{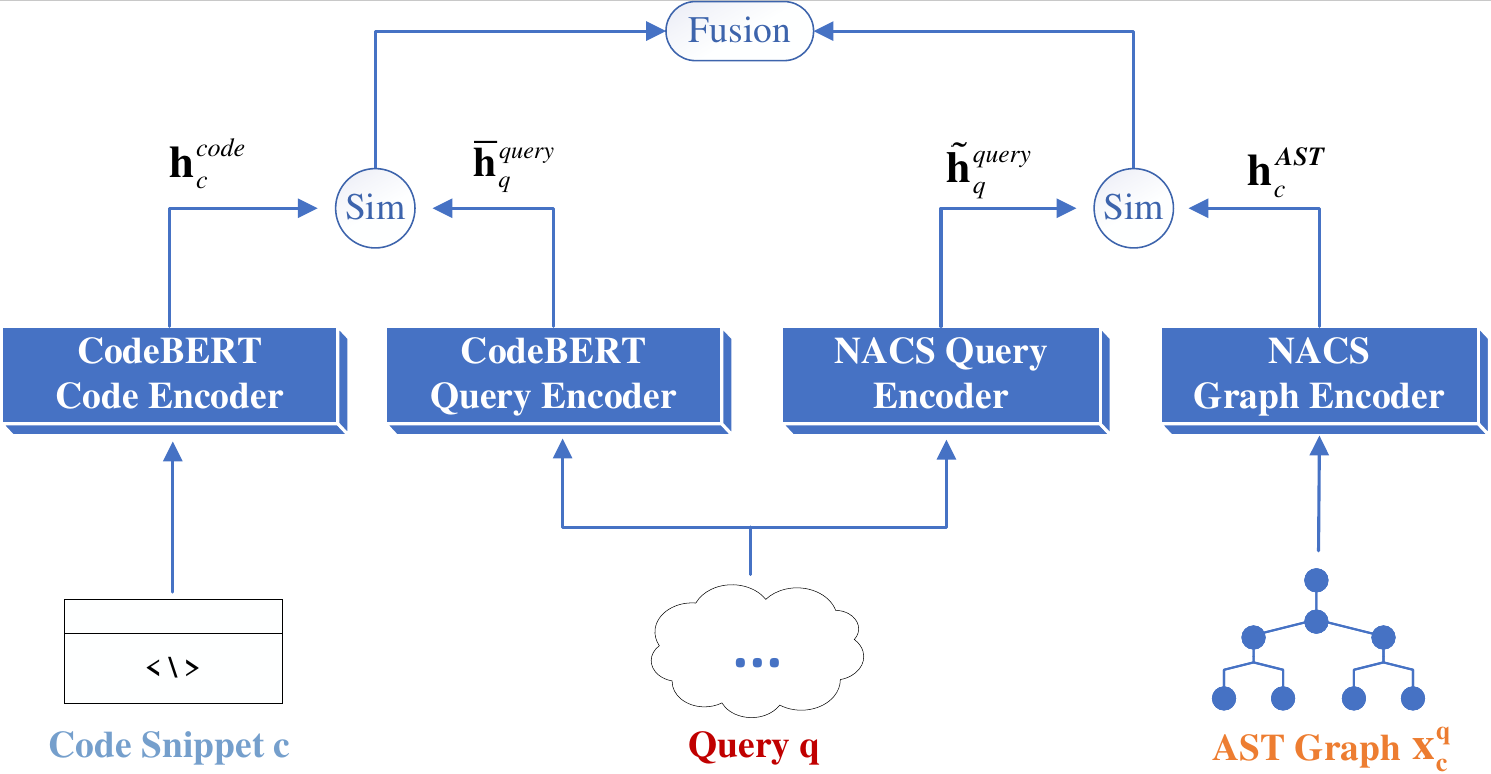}
    \caption{Fine-tune \ours.}
    \label{fig:enhance}
\end{figure}

\subsubsection{Derive Query Representations}
\label{sec:queryrep}

Each query $q$ is fed into the pre-trained CodeBERT~\citep{FengGTDFGS0LJZ20} and the query encoder of the pre-trained \ours to derive two query representations, respectively: 
\begin{equation}
\label{eq:query_rep}
\begin{aligned}
\mathbf{\bar{h}}_{q}^{query} &= \textit{CodeBERT}_{query}(q)\\
\mathbf{\tilde{h}}_{q}^{query} &= \textit{NACS}_{query}(q)
\end{aligned}
\end{equation}

As depicted in Figure~\ref{fig:enhance}, we calculate two matching scores for the query $q$ and each code snippet $c$ in the codebase:
\begin{equation}
\begin{aligned}
score_1{(q,c)} &= \textit{sim}(\mathbf{h}_{c}^{code},\,\, \mathbf{\bar{h}}_{q}^{query})\\
score_2{(q,c)} &= \textit{sim}(\mathbf{h}_{c}^{AST},\,\, \mathbf{\tilde{h}}_{q}^{query})
\end{aligned}
\end{equation}
where $score_1{(q,c)}$ is the matching score from CodeBert and $score_2{(q,c)}$ indicates the matching score considering the multi-view information of $x_c^q$.

\subsubsection{Fine-Tune \ours}

Given code and query representations from the training data, we fine-tune CodeBert and \ours to enhance code search with multi-view representations. 
Figure~\ref{fig:enhance} depicts this process.
$score_1{(q,c)}$ and $score_2{(q,c)}$ is fused to indicate the matching score between a query $q$ and a code snippet $c$:
\begin{equation}
\label{eq:rel_score}
score = score_1{(q,c)} + \lambda \cdot score_2{(q,c)},
\end{equation}
where $\lambda$ is a pre-defined parameter for balancing two parts.

During optimization, given a mini-batch of $N$ query-code training pairs, the overall framework is trained to minimize the following loss:
\begin{equation}\label{eq:cs}
\mathcal{L}_{cs}=- \log \frac{e^ {score^{(i,i)}/ \tau}}{\sum_{v=0}^{N} e^{ score^{(i,j)}/ \tau}}.
\end{equation}

\subsubsection{Search Relevant Code Snippets with \ours}

At search time, given an input query $q$, the relevance scores between $q$ and each code snippet $c$ in the codebase is computed as Equation~\ref{eq:rel_score}.
Relevant scores between $q$ and all candidate code snippets are sorted and the top-K code snippets with highest relevance scores are returned as the search result for the query $q$.

This step is light-weight: the representations of all code snippets in the codebase can be pre-computed, and the two query representations (Equation~\ref{eq:query_rep}) can be directly retrieved from the fine-tuned CodeBert and \ours.

\section{Experiments}
\label{sec:exp}

In this section, we will report and analyze our experimental results to answer the following research questions:
\begin{itemize}
  \item \textbf{RQ1}: How effective is \ours compared to state-of-the-art code search methods for the code search task? Is \ours able to achieve naming-agnostic code search? (Section~\ref{sec:overallper})
  \item \textbf{RQ2}: How does each component in \ours contribute to the code search performance? (Section~\ref{sec:contri})
  \item \textbf{RQ3}: Can \ours be adopted to improve existing code search methods? (Section~\ref{sec:generalization})
  \item \textbf{RQ4}: How do different hyper-parameter values affect the search result? (Section~\ref{sec:parameter})
\end{itemize}

\subsection{Experiment Settings}

\subsubsection{Data}
We choose four public datasets that are commonly used in previous works in our experiments:
\begin{enumerate}
\item \textbf{CodeSearchNet-Python dataset}\footnote{\url{https://github.com/github/CodeSearchNet}}~\citep{abs-1909-09436}: It is the Python dataset used in the CodeSearchNet challenge.
It contains 0.5M Python functions and their corresponding natural language descriptions.
CodeSearchNet dataset is widely used in various code relevant tasks including but not limited to code representation learning~\citep{FengGTDFGS0LJZ20,GuoRLFT0ZDSFTDC21}, code summarization~\citep{LinOZCLW21} and code completion~\citep{CiniselliCPPPB21}.

\item \textbf{CodeSearchNet-Java dataset}\footnote{\url{https://huggingface.co/datasets/Nan-Do/code-search-net-java}}~\citep{abs-1909-09436}: It is the Java dataset used in the CodeSearchNet challenge.
It contains 0.5M Java functions and their corresponding natural language descriptions.

\item \textbf{CoSQA dataset}\footnote{\url{https://github.com/microsoft/CodeXGLUE/tree/main/Text-Code/NL-code-search-WebQuery}}~\citep{HuangTSG0J0D20}: It is the dataset released by Microsoft. It contains 20,604 labeled query-code pairs. Each query is written in English while each code snippet is a Python code snippet. The data is annotated by at least 3 human annotators. The queries come from the search logs of the Microsoft Bing search engine, and each code snippet is a Python function crawled from GitHub.

\item \textbf{CoSQA-Var dataset}: To investigate the performance of different code search models and \ours when handling the naming issue, we randomly replace variable names in CoSQA dataset and construct a new dataset CoSQA-Var. For a variable in a code snippet of CoSQA, we randomly replace its variable name with another variable name used in CoSQA. The random replacement keeps the consistency of variable names: after the replacement, a variable will still use the same (new) variable name within one code snippet. The CoSQA-Var contains the same number of query-code pairs as the CoSQA dataset.

\end{enumerate}

\subsubsection{Evaluation Metric}

To evaluate the performance, we adopt Mean Reciprocal Rank (MRR), the most widely used evaluation metric for code search.
MRR is the average of the reciprocal ranks of results for test queries:
\begin{equation}
\label{eq:mrr}
\text{MRR} = \frac{1}{\vert Q \vert} \sum_{i=1}^{\vert Q \vert} \frac{1}{rank_i},
\end{equation}
where $\vert Q \vert$ is the number of queries and $rank_i$ indicates the rank of the ground-truth code snippet w.r.t. the $i$-th query.

\subsubsection{Baselines}

We use seven prevalent code search models as baselines in our experiments:
\begin{enumerate}
    \item \textbf{DeepCS}\footnote{\url{https://github.com/guxd/deep-code-search}}~\citep{GuZ018}: DeepCS is the pioneering code search method that adopts deep learning. DeepCS considers three aspects of source code: the method name, the API invocation sequence, and the tokens contained in the source code. The camel split tokens of methods names and the API invocation sequences are embedded by two RNNs, respectively. Code tokens are embedded via a multilayer perceptron (MLP). The three representation vectors are fused into the code representation vector via a fully connected layer. Query tokens are embedded using an RNN. Finally. the matching score between a query and a code snippet is measured by the cosine similarity between the code representation vector and the query representation vector.

    \item \textbf{OCoR}\footnote{\url{https://github.com/pkuzqh/OCoR}}~\citep{ZhuSLXZ20}: OCoR is proposed with a specific focus on the overlaps between identifiers in code and words in queries. OCoR uses two specifically-designed components to capture overlaps: the first embeds names by characters to capture the overlaps between names, and the second introduces an overlap matrix to represent the degrees of overlaps between each natural language word and each identifier.

    \item \textbf{CSRS}\footnote{\url{https://github.com/css518/CSRS}}~\citep{ChengK22}: CSRS considers both relevance matching and semantic matching for code search. It contains a relevance matching module that measures lexical matching signals between the query and the code snippets, and a co-attention based semantic matching module to capture the semantic correlation between the query and the code snippets.

    \item \textbf{CoCLR}\footnote{\url{https://github.com/Jun-jie-Huang/CoCLR}}~\citep{HuangTSG0J0D20}: CoCLR adopts CodeBERT to encode queries and code snippets into representations that are further fed into a MLP to measure query-code relevance.

    \item \textbf{CodeBERT}\footnote{\url{https://github.com/microsoft/CodeBERT}\label{footnote:codebert}}~\citep{FengGTDFGS0LJZ20}: CodeBERT is a bimodal pre-trained model for programming language and natural language. CodeBERT shares the same architecture as RoBERTa~\citep{abs-1907-11692} and it is pre-trained via the masked language modeling task~\cite{DevlinCLT19} and the replaced token detection task on the CodeSearchNet dataset.

    \item \textbf{GraphCodeBERT}$\textsuperscript{\ref{footnote:codebert}}$~\citep{GuoRLFT0ZDSFTDC21}: GraphCodeBERT is a pre-trained model for programming language and natural language. GraphCodeBERT adopts the idea of BERT~\cite{DevlinCLT19} and uses data flow which is the semantic-level structure that encoders the data flow relations between variables. It is pre-trained via the masked language modeling task, the code structure prediction task and the alignment task between code and structure on the CodeSearchNet dataset. 

    \item \textbf{DGMS}\footnote{\url{https://github.com/ryderling/DGMS}}~\citep{LingWWPMXLWJ21}: DGMS is an end-to-end deep graph matching and searching model based on graph neural networks for code search. DGMS represents both queries and code snippets with the unified graph-structured data, and then use the proposed graph matching and searching model to retrieve the best matching code snippet.

\end{enumerate}

\subsubsection{Implementation Details}

For Python datasets CodeSearchNet-Java, CoSQA and CoSQA-Var, we extract ASTs and conduct data augmentation for \ours using Python official AST library\footnote{\url{https://docs.python.org/3/library/ast.html}}.
For Java dataset CodeSearchNet-Java, we extract ASTs and conduct data augmentation for \ours using JavaParser library\footnote{\url{https://javaparser.org}}.
We use the implementations from authors for baselines.
We implement \ours using PyTorch.
The implementation and datasets are available at \url{https://github.com/KDEGroup/NACS}.

The experiments were run on a machine with two Intel(R) Xeon(R) CPU E5-2678 v3 @ 2.50 GHz, 256 GB main memory and 8 GeForce RTX 2080 Ti graphics cards with 11 GB memory
per card. 
During running, each program will monopolize one graphics card even if it does not
require the complete 11 GB video memory.
Performance is reported based on the average of five runs of models.

We set batch size to 48. 
The dimensions of AST path node embedding, AST path embedding and query embedding are 128.
The dimensions of AST graph node embedding and AST graph embedding are 128.
Adam optimizer~\citep{KingmaB14} is used with a initial learning rate 0.001.
\ours and baselines use same settings whenever it is possible.
Otherwise, the default settings for baselines are adopted to achieve a fair comparison.
We conduct a grid search to find best settings for $\lambda$ and $\tau$ for \ours. 
We find that $\lambda=0.0001$ and $\tau=0.07$ bring best results and we use them as the default setting. 
In Section~\ref{sec:parameter}, we report the impacts of different values of $\lambda$ and $\tau$ on \ours.

\subsection{Analysis of Code Search Performance (RQ1)}
\label{sec:overallper}

\subsubsection{Overall Performance}

\begin{table}[t]
\centering
\caption{MRR scores of each method on the four datasets.}
\label{tab:result1}
\scalebox{0.9}{
\begin{tabular}{|c|c|c|c|c|}
\hline
Model         & CodeSearchNet-Python & CodeSearchNet-Java & CoSQA & CoSQA-Var \\ \hline
OCoR          & 0.165                & 0.201              & 0.352 & 0.293     \\ \hline
CSRS          & 0.231                & 0.386              & 0.373 & 0.231     \\ \hline
DeepCS        & 0.164                & 0.199              & 0.472 & 0.433     \\ \hline
CodeBERT      & 0.665                & 0.667              & 0.652 & 0.558     \\ \hline
GraphCodeBERT & 0.694                & 0.689              & 0.648 & 0.456     \\ \hline
CoCLR         & 0.637                & 0.631              & 0.647 & 0.515     \\ \hline
DGMS          & 0.579                & 0.555              & 0.471 & 0.399     \\ \hline
\ours         & 0.701                & 0.705              & 0.708 & 0.704     \\ \hline
\end{tabular}
}
\end{table}

Firstly, we investigate the overall code search performance of all methods on CodeSearchNet-Python, CodeSearchNet-Java, CoSQA and CoSQA-Var datasets.
Table~\ref{tab:result1} shows the MRR scores of each code search model. 
From the results, we have the following observations:
\begin{enumerate}
  \item Compared to OCoR, CSRS, DeepCS and DGMS, pre-training based methods CodeBERT, GraphCodeBERT, CoCLR and \ours achieve much better performance on all the four datasets, showing the effectiveness of the Bert-style pre-training method on improving code search performance.

  \item Compared to all baselines, \ours achieves highest MRR scores and it consistently outperforms the best two baselines GraphCodeBert and CoCLR on the four datasets, showing that the robustness of \ours.

  \item The CoSQA-Var dataset is designed for evaluating the performance of code search models when variables with same meaning have different variable names in different code snippets. From Table~\ref{tab:result1}, we can see that the performance of all baselines significantly degrades on the CoSQA-Var dataset ($\downarrow$ 8.26\%-38.07\%) compared to their performance on the CoSQA dataset. Differently, the performance of \ours is not affected by changing variable names in code snippets. This observation has verified the effectiveness of our proposed contrastive multi-view code representation learning: capturing the structural information in ASTs without explicitly binding code tokens to specific pre-trained representations can help overcome the problem of variable naming when modeling code snippets.

  \item \ours performs similarly on both CodeSearchNet-Python and CodeSearchNet-Java and it consistently outperforms baselines, showing the effectiveness of \ours on both Python and Java.

\end{enumerate}

In summary, \ours consistently shows superior performance than all baselines.
The gap is more significant on CoSQA-Var, showing the effectiveness of \ours on handling the naming issue.

\begin{table}[t]
\centering
\caption{Running time of each method on CodeSearchNet-Java and CoSQA.}
\label{tab:resulttime}
\scalebox{0.9}{
\begin{tabular}{|c|rr|rr|}
\hline
\multirow{2}{*}{Model} & \multicolumn{2}{c|}{CodeSearchNet-Java}                        & \multicolumn{2}{c|}{CoSQA}                                    \\ \cline{2-5} 
                       & \multicolumn{1}{c|}{Train}         & \multicolumn{1}{c|}{Test} & \multicolumn{1}{c|}{Train}        & \multicolumn{1}{c|}{Test} \\ \hline
OCoR                   & \multicolumn{1}{r|}{284.968 hours} & 4.017 hours               & \multicolumn{1}{r|}{742.260 min.} & 261.901 sec.              \\ \hline
CSRS                   & \multicolumn{1}{r|}{14.638 hours}  & 6.445 hours               & \multicolumn{1}{r|}{37.254 min.}  & 432.371 sec.              \\ \hline
DeepCS                 & \multicolumn{1}{r|}{75.762 hours}  & 10.932 hours              & \multicolumn{1}{r|}{195.437 min.} & 737.075 sec.              \\ \hline
CodeBERT               & \multicolumn{1}{r|}{15.962 hours}  & 2028.199 sec.            & \multicolumn{1}{r|}{41.611 min.}  & 8.815 sec.                \\ \hline
GraphCodeBERT          & \multicolumn{1}{r|}{33.977 hours}  & 284.899 sec.              & \multicolumn{1}{r|}{87.956 min.}  & 10.715 sec.               \\ \hline
CoCLR                  & \multicolumn{1}{r|}{55.364 hours}  & 2583.092 sec.             & \multicolumn{1}{r|}{109.104 min.} & 46.374 sec.               \\ \hline
DGMS                   & \multicolumn{1}{r|}{45.191 hours}  & 1553.489 sec.             & \multicolumn{1}{r|}{115.999 min.} & 12.742 sec.               \\ \hline
\ours                  & \multicolumn{1}{r|}{18.740 hours}  & 531.692 sec.              & \multicolumn{1}{r|}{48.529 min.}  & 34.658 sec.               \\ \hline
\end{tabular}
}
\end{table}

\subsubsection{Running Time}
\label{sec:runningtime}

We report the running time of each methods in Table~\ref{tab:resulttime}.
Since CodeSearchNet-Python and CodeSearchNet-Java have similar amounts of data and CoSQA and CoSQA-Var have the same amount of data, we only provide the training and the test time on CodeSearchNet-Java and CoSQA in Table~\ref{tab:resulttime}.
From the results, we have the following observations:
\begin{enumerate}
  \item OCoR has the longest running time. The reason is that, during training, it counts the overlaps between identifiers in code and words in queries, which is a costly operation.

  \item Pre-training based methods CodeBERT, GraphCodeBERT, CoCLR and \ours do not require much more training time compared to small models CSRS and DGMS. And their training time is even shorter in some cases. We can also observe that the test time of pre-training methods are generally much shorter compared to OCoR, CSRS, DeepCS and DGMS, showing that pre-training methods are efficient during inference and do not affect user experience on code search.

  \item The running time of \ours is at the same level as the running time of other pre-training methods. On the largest dataset CodeSearchNet-Java (it is 25 times larger than CoSQA), its running time is acceptable compared to other baselines, showing the scalability of \ours on large codebases.

\end{enumerate}

Overall, the running overhead of \ours is at the same level as other pre-training based methods and it shows strong scalability on large codebase.

\subsubsection{Case Studies of Code Search Results}
\label{sec:casestudy}

We provide two types of case studies of code search results to further illustrate that the effectiveness of \ours on handling same variables with different variable names (i.e., naming-agnostic code search).

\begin{table*}[t]
\centering
\caption{Case Study 1: Ranking results of ground-truth code snippets w.r.t. two queries using \ours, CoCLR, GraphCodeBERT, DeepCS and DGMS on both CoSQA and CoSQA-Var datasets.}
\label{tab:casestudy1}
\resizebox{\textwidth}{!}{
\begin{tabular}{c}
\includegraphics{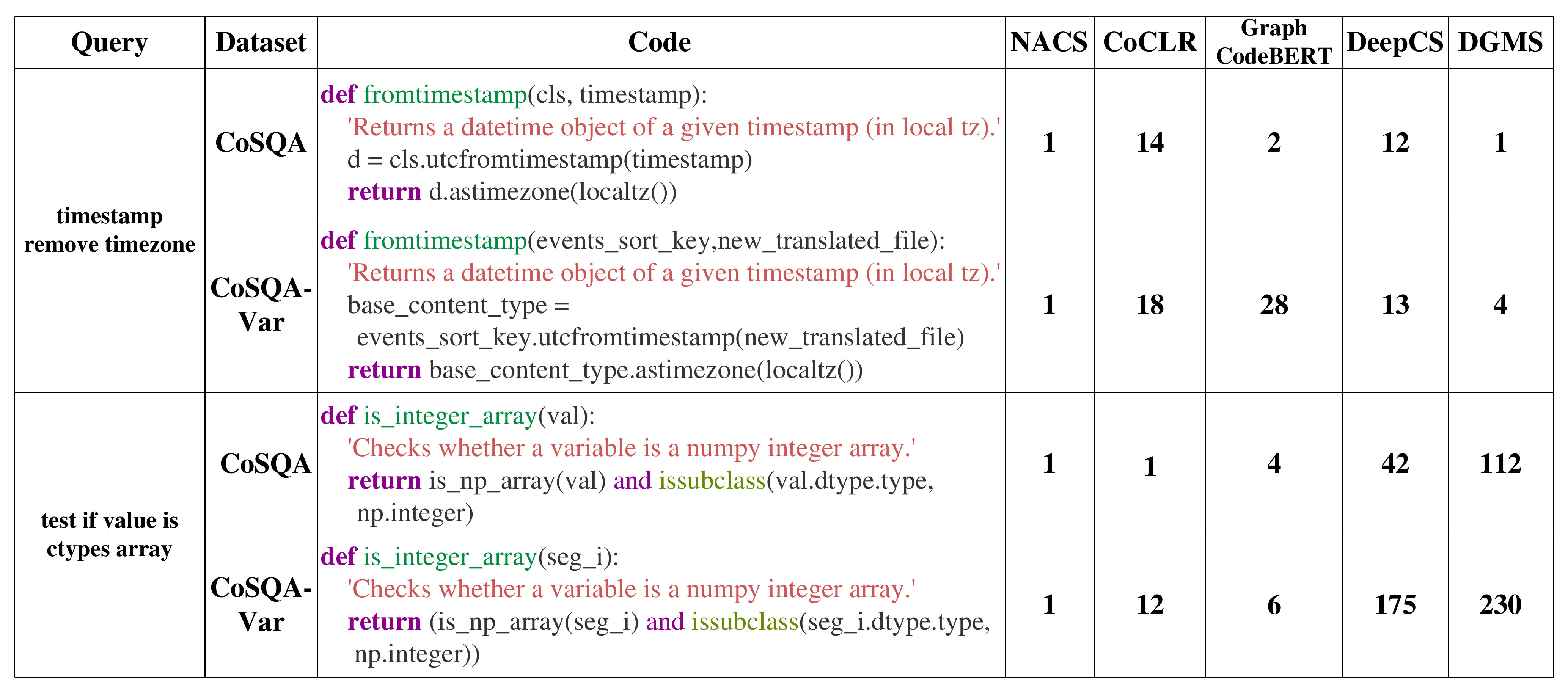}
\end{tabular}
}
\end{table*}

Table~\ref{tab:casestudy1} illustrates the ranking results of ground-truth code snippets w.r.t. two queries using \ours, CoCLR, GraphCodeBERT, DeepCS and DGMS on both CoSQA and CoSQA-Var datasets.
We show the ranking position of the ground-truth code snippet using each method.
From the results, we can observe that: 
\begin{enumerate}
\item The accuracy of CoCLR, GraphCodeBert, DeepCS and DGMS are affected when variable names are randomly replaced and they provide worse results for the same query on the CoSQA-Var dataset compared to their original search results on the CoSQA dataset. Differently, \ours provides consistent ranking results for the ground-truth code snippet w.r.t. the same query on the two datasets, showing that the performance of \ours is not affected by the change of variable names. In other words, \ours is able to achieve naming-agnostic code search.

\item For cases where different code search methods have comparable performance (e.g., the second case in Table~\ref{tab:casestudy1} where both \ours and CoCLR rank the ground-truth code snippet as the top-1 search result on the CoSQA dataset), randomly replacing variable names in the data can be used to further identify the performance gap between different methods and the deficiency of existing methods (e.g., the performance difference between \ours and CoCLR on the CoSQA-Var dataset is distinguishable in the second case of Table~\ref{tab:casestudy1}).
\end{enumerate}

\begin{table*}[t]
\centering
\caption{Case Study 2: Ranking results of ground-truth code snippets w.r.t. two queries using \ours, CoCLR, GraphCodeBERT, DeepCS and DGMS on the CoSQA dataset. Ground-truth code snippets are highlighted in yellow.}
\label{tab:casestudy2}
\resizebox{\textwidth}{!}{
\begin{tabular}{c}
\includegraphics{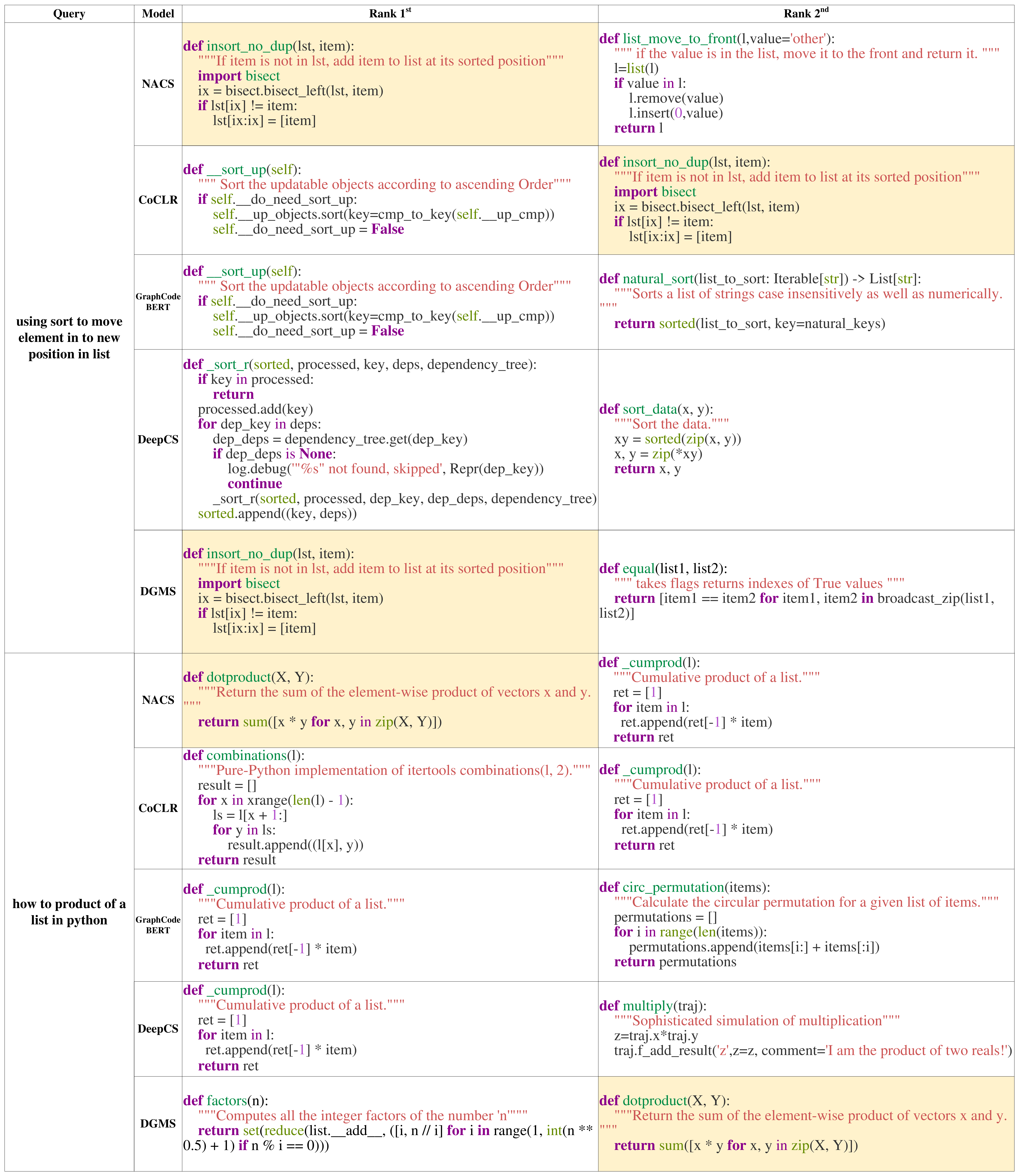}
\end{tabular}
}
\end{table*}

Table~\ref{tab:casestudy2} depicts the ranking results of ground-truth code snippets w.r.t. two queries using \ours, CoCLR, GraphCodeBERT, DeepCS and DGMS on the CoSQA dataset.
We show the top-2 retrieved code snippet results from each method and the ground-truth code snippet is shown in bold.
We can see that: 
\begin{enumerate}
\item \ours precisely ranks ground-truth code snippets first, while the other four baselines cannot consistently rank ground-truth code snippets as the top-$1$ results. 
The observation illustrates that \ours exceeds existing code search methods in general cases.

\item Moreover, top-1 results provided by CoCLR, GraphCodeBERT and DeepCS for the first case in Table~\ref{tab:casestudy2} all contain the code token ``sort'', which is part of the input query.
The ground-truth code snippet does not contain the code token ``sort'', which is possibly a reason why the three baselines cannot correctly rank the ground-truth code snippet.
Hence, the three baselines emphasize on the overlap between code snippets and queries, meaning that they may only capture the token-level information and cannot fully understand the deep semantics of code snippets.
Differently, \ours can correctly find the ground-truth code snippet even if it does not contain code tokens that appear in queries.
\end{enumerate}

\subsubsection{Relations between Learned Query Space and Code Space}
\label{sec:vis}

\begin{figure}[t]
    \centering
    \includegraphics[width=0.55\textwidth]{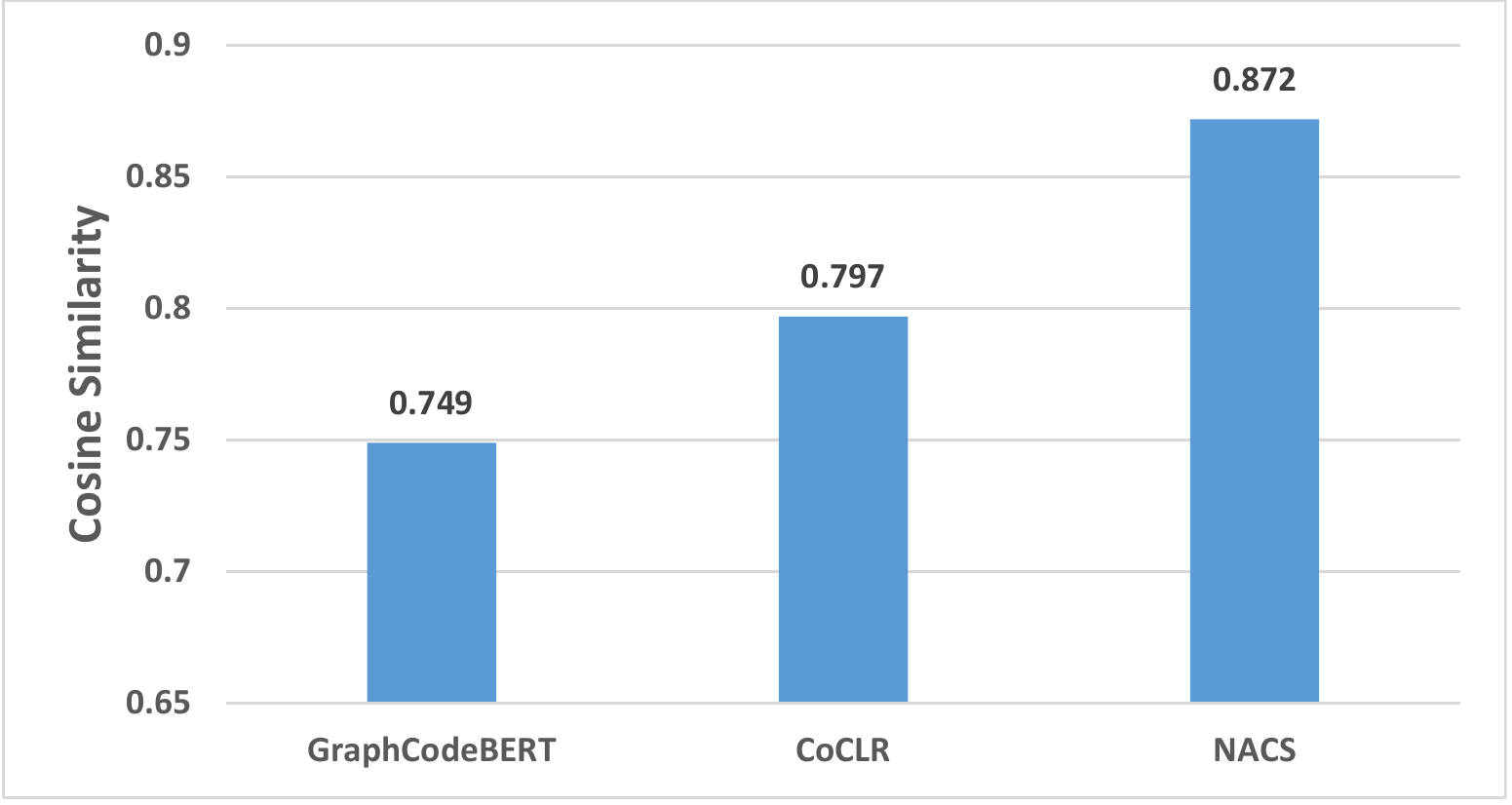}
    \caption{Cosine similarity between query and corresponding ground-truth code snippet on the CoSQA dataset.}
    \label{fig:cosine}
\end{figure}

Code search models aim at matching a query and the ground-truth code snippet. 
The similarity between the query embedding and the code embedding can somehow demonstrate the ability of a code search model to match queries and corresponding ground-truth code snippets.
We use the average embeddings ($\mathbf{t}$) of tokens in a query to represent the query and use the average embeddings ($\mathbf{c}$) of code tokens in the corresponding ground-truth code snippet to indicate the ground-truth code snippet.
We calculate the average cosine similarity between $\mathbf{t}$ and $\mathbf{c}$ and visualize the results of GraphCodeBERT, CoCLR and \ours on CoSQA in Figure~\ref{fig:cosine}.
From Figure~\ref{fig:cosine}, we can see that the average similarity between $\mathbf{t}$ and $\mathbf{c}$ for \ours is highest, showing that \ours is better at matching a query and the ground-truth code snippet.

\subsection{Contributions of Each Component in \ours (RQ2)}
\label{sec:contri}

To investigate the impact of each component in \ours, we conduct an ablation study to analyze whether each major component in \ours contribute to the overall performance of \ours.
We have designed several variations of \ours and tested them on the CoSQA dataset:
\begin{itemize}
\item \textbf{\oursgraph}: This variation does not consider contrasting graph view of ASTs shown in Equation~\ref{eq:graphviewloss1}, and the pre-training loss in Equation~\ref{eq:loss} does not include $\mathcal{L}_g$.

\item \textbf{\ourspath}: This variation does not adopt the cross-view contrast (i.e., contrast path view and graph view) shown in Equation~\ref{eq:pathviewloss1}, and the pre-training loss in Equation~\ref{eq:loss} does not contain $\mathcal{L}_p$.

\item \textbf{\oursast}: This variation removes all AST-related components (graph-view contrast and cross-view contrast) and it is degraded to CodeBERT.

\end{itemize}

Table~\ref{tab:ablation-study} reports the result of the ablation study.
From the result, we can see that:
\begin{enumerate}

  \item \oursast (i.e., CodeBERT) shows the worst performance among all \ours variations, which indicates that the superior performance of \ours comes from adopting the naming-agnostic contrastive multi-view code representation learning instead of CodeBERT.

  \item \oursgraph and \ourspath show better code search performance than \oursast, showing that using graph-view contrast (\oursgraph) or cross-view contrast (\ourspath) alone can improve the quality of code search. In other words, contrastive pre-training based on the structural information of ASTs can help enhance the code search model.

  \item \ours shows even better performance than \oursgraph and \ourspath, demonstrating that graph-view contrast and cross-view contrast can be used together to help the code search model alleviate the impact of different naming styles and achieve high-quality code search results.

\end{enumerate}

\subsection{Effectiveness of \ours on Improving Other Code Search Models (RQ3)}
\label{sec:generalization}

\ours can be easily adopted to improve existing code search models: we can adopt any existing deep learning based code search model to generate representations for candidate code snippets and queries as a replacement of CodeBERT used in Section~\ref{sec:coderep} and Section~\ref{sec:queryrep}.

Table~\ref{tab:others} reports the results of using \ours to enhance DeepCS, CoCLR and GraphCodeBERT.
In Table~\ref{tab:others}, the superscript ``*'' indicates the corresponding method is enhanced by \ours.
By comparing results of one code search method and its improved version in Table~\ref{tab:others}, we can find that \ours indeed improves the search performance of existing methods.
Furthermore, the performance of DeepCS, GraphCodeBert and CoCLR on CoSQA-Var is much worse than their performance on CoSQA, and the decline percentages are 8.26\%, 42.11\% and 20.40\%, showing that existing code search methods suffer from the naming issue.
As a comparison, the performance decline is unremarkable for DeepCS$^*$ (3.07\%), GraphCodeBert$^*$ (2.32\%) and CoCLR$^*$ (2.95\%). 
Therefore, we can conclude that \ours can be adopted to improve the robustness of existing code search methods and help them alleviate the negative impact of  different naming styles used in code snippets.

\begin{table}[t]
\centering
\caption{Performance of different variations of \ours on the CoSQA dataset.}
\label{tab:ablation-study}
\setlength{\tabcolsep}{10mm}{
\begin{tabular}{|c|c|}
\hline
Variations    & MRR Score  \\ \hline
\ours         & 0.708      \\ \hline
\oursgraph    & 0.682      \\ \hline
\ourspath     & 0.687      \\ \hline
\oursast      & 0.652      \\ \hline
\end{tabular}}
\end{table}

\begin{table}[t]
\centering
\caption{Performance of existing code search methods enhanced by \ours.}
\label{tab:others}
\begin{tabular}{|l|l|l|l|}
\hline
\textbf{Method}     & \multicolumn{1}{l|}{\textbf{CoSQA}} & \multicolumn{1}{l|}{\textbf{CoSQA-Var}} & \textbf{Decline Percentage} \\ \hline\hline
DeepCS              & 0.472                               & 0.433                                   & $\downarrow$8.26\%          \\ \hline
DeepCS$^{*}$        & 0.488                               & 0.473                                   & $\downarrow$3.07\%          \\ \hline\hline
GraphCodeBERT       & 0.648                               & 0.456                                   & $\downarrow$42.11\%         \\ \hline
GraphCodeBERT$^{*}$ & 0.691                               & 0.675                                   & $\downarrow$2.32\%          \\ \hline\hline
CoCLR               & 0.647                               & 0.515                                   & $\downarrow$20.40\%         \\ \hline
CoCLR$^{*}$         & 0.677                               & 0.657                                   & $\downarrow$2.95\%          \\ \hline
\end{tabular}
\end{table}

\subsection{Impacts of Hyper-Parameters on Search Results (RQ4)}
\label{sec:parameter}

\begin{figure}[t]
    \centering
    \includegraphics[width=0.75\textwidth]{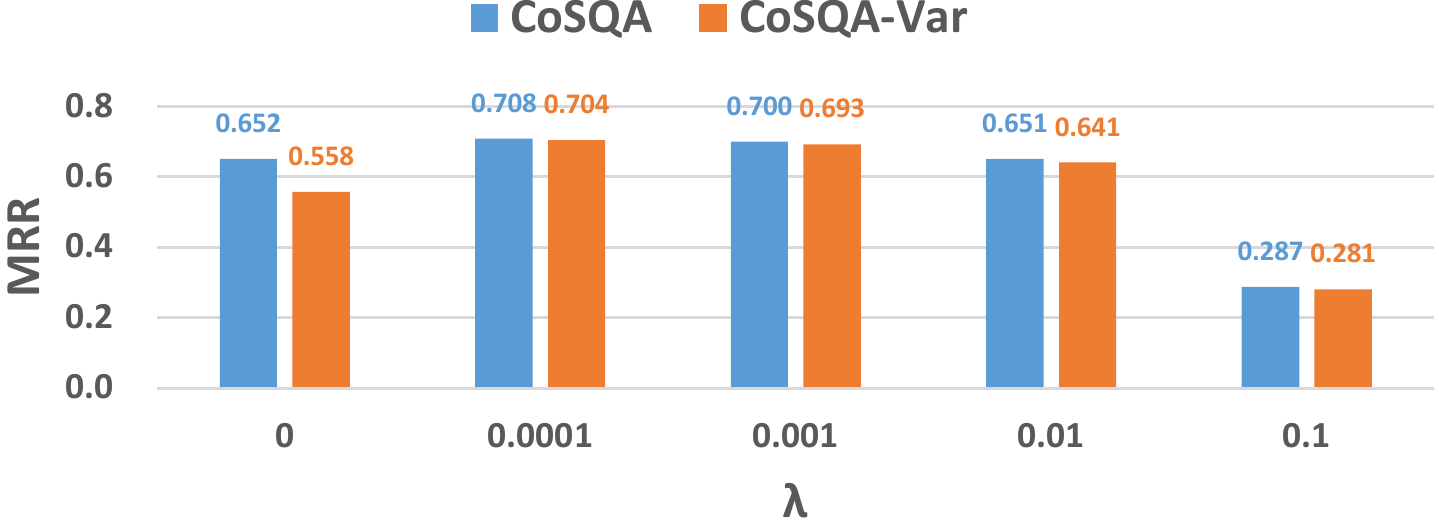}
    \caption{Impacts of $\lambda$ on the performance of \ours.}
    \label{fig:lambda}
\end{figure}

\begin{figure}[t]
    \centering
    \includegraphics[width=0.75\textwidth]{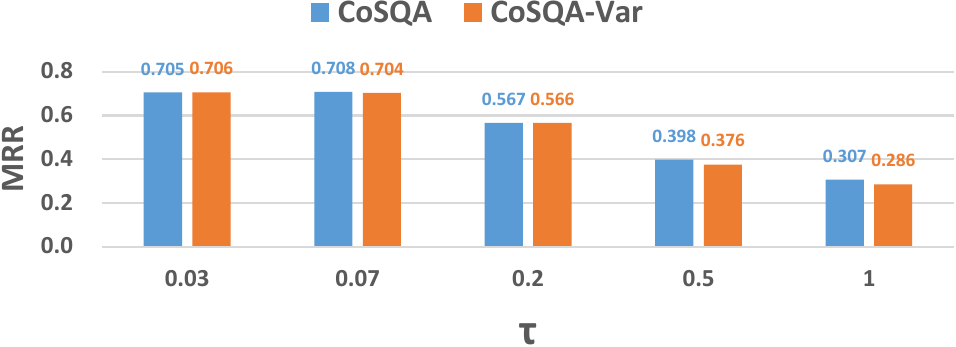}
    \caption{Impacts of $\tau$ on the performance of \ours.}
    \label{fig:tau}
\end{figure}

We investigate the impacts of $\lambda$ in Equation~\ref{eq:rel_score} on the performance of \ours.
Figure~\ref{fig:lambda} illustrates the performance of \ours on CoSQA and CoSQA-Var datasets when $\lambda$ is set to 0, 0.0001, 0.001, 0.01 and 0.1, respectively.
We can observe that the best result are achieved when $\lambda$ is set to 0.0001 and larger $\lambda$ ($>0.1$) will significantly degrade the performance of \ours.
Overall, it is easy to tune $\lambda$ as the best value falls in a relative narrow range.
When $\lambda$ is set to 0, \ours is identical to CodeBERT. 
The performance of \ours declines from 0.708/0.704 to 0.652/0.558 when $\lambda$ is changed from 0.0001 to 0.
The noticeable performance reduction shows that our proposed naming-agnostic contrastive multi-view code representation learning indeed improve the performance of code search.

We further report the influence of the temperature parameter $\tau$. 
Figure~\ref{fig:tau} provides the performance of \ours on CoSQA and CoSQA-Var datasets when varying $\tau$ in 0.03, 0.07, 0.2, 0.5 and 1.
Best performance is achieved when setting $\tau$ to 0.03 (for CoSQA-Var) and 0.07 (for CoSQA).
As $\tau$ continues to increase, the performance of \ours declines.
The impacts of $\tau$ on \ours show it is easy to tune and find a suitable setting for $\tau$.

Considering the results in Figure~\ref{fig:lambda} and Figure~\ref{fig:tau}, we can again confirm our conclusion obtained in Section~\ref{sec:overallper}: \ours is almost invariable to different naming conventions since its performance on CoSQA and CoSQA-Var datasets, when using same hyper-parameters, are close.

\section{Conclusion}
\label{sec:con}

Code search provides a way for developers to reuse implementations and learn new API usage, reducing developers' cognitive burden.
Due to the pivotal role of code search in software development, much effort has been devoted to improving the quality of code search.
Particularly, there is a surge of works on deploying deep learning techniques to enhance code search.
Nevertheless, most existing deep learning based code search methods ignore the impact of different naming conventions, causing downgraded performance when the same variable has different variable names in different code snippets.

In this paper, we propose a naming-agnostic code search method \ours by using our designed contrastive multi-view code representation learning.
\ours strips information bound to variable names from AST and focuses on capturing intrinsic properties solely from AST structures. 
We use semantic-level and syntax-level augmentation techniques to prepare realistically rational data and adopt contrastive learning to design a graph-view modeling component in \ours to enhance the understanding of code snippets. 
ASTs are further modeled from a path view to strengthen the graph-view modeling component via multi-view learning. 
Extensive experiments show that \ours outperforms baselines and it can also be adapted to help existing code search methods overcome the impact of different naming conventions.

In the future, we plan to explore other designs of lightweight architectures to reduce both the training code and the search cost of \ours to reduce the cost of code search and further improve the scalability of \ours. We also plan to collect more code search data containing real queries and corresponding code snippets written in more programming languages to construct large-scale evaluation benchmarks for better evaluating the code search task.

\begin{acks}
This work was partially supported by Natural Science Foundation of Xiamen, China (No. 3502Z202471028), National Natural Science Foundation of China (No. 62002303, 42171456) and CCF-Tencent Open Fund (RAGR20210129).
\end{acks}

\bibliographystyle{ACM-Reference-Format}
\bibliography{main}

\end{document}